\documentclass[preprint, aps, prd, preprintnumbers, superscriptaddress, nofootinbib]{revtex4-1}

\usepackage{amsmath}
\usepackage{bm}
\usepackage{amsfonts}
\usepackage{graphicx}
\usepackage{hyperref}
\usepackage{color}

\hypersetup{
colorlinks=true,
linkcolor=blue,
linktoc=page,
citecolor=blue,
urlcolor=blue}

\begin{document}

\title{\bf The Swampland Conjecture Bound Conjecture}

\newcommand{\FIRSTAFF}{\affiliation{Department of Physics, University at Buffalo, Buffalo, NY 14260, USA}}

\author{William H. Kinney}
\email[Electronic address: ]{@WKCosmo}
\FIRSTAFF

\date{April 1, 2021}
\begin{abstract}
I conjecture an upper bound on the number of possible swampland conjectures by comparing the entropy required by the conjectures themselves to the Beckenstein-Hawking  entropy of the cosmological horizon. Assuming of order 100 kilobits of entropy per conjecture, this places an upper bound of order $10^{117}$ on the number of conjectures.  I estimate the rate of production of swampland conjectures by the number of papers listed on INSPIRE with the word ``swampland'' in the title or abstract, which has been showing approximately exponential growth since 2014. At the current rate of growth, the entropy bound on the number of swampland conjectures can be projected to be saturated on a timescale of order $10^{-8} H_0^{-1}$. I compare the upper bound from the Swampland Conjecture Bound Conjecture (SCBC) to the estimated number of vacua in the string landscape. Employing the duality suggested by AdS/CFT between the quantum complexity of a holographic state and the volume of a Wheeler-Dewitt spacetime patch, I place a conservative lower bound of order $\mathcal{N}_H > 10^{263}$ on the number of Hubble volumes in the multiverse which must be driven to heat death to fully explore the string landscape via conjectural methods. 
\end{abstract}

\maketitle

\section{Introduction}

The so-called ``Swampland'' program has, in the past decade, revolutionized our conception of quantum gravity. Or something. The existence of an inconceivably vast ``Landscape'' of stable vacua of string theory has long been conjectured, for example by Susskind, who suggests that ``whether we like it or not, this is the kind of behavior that gives credence to the Anthropic Principle'' \cite{Susskind:2003kw}.\footnote{Anything Susskind says is by defnition true.} An alternate approach to investigating the properties of quantum gravity was pioneered by Vafa \cite{Vafa:2005ui}, applying the method of \textit{calculatus eliminatus} \cite{Cat:1971meow} to formulate conjectures describing what a UV-complete effective theory is \textit{not}. The string ``Swampland'' is accordingly defined to be the set of low-energy effective field theories which are incompatible with a consistent UV-complete limit. The best-known example of this is the Weak Gravity Conjecture \cite{ArkaniHamed:2006dz}, which postulates that any UV-complete theory must possess at least one light state carrying a $U(1)$ charge. Otherwise, the Hawking radiation would result in the presence of stable extremal Reissner-Nordstrom black holes. Which are bad. Other more recent conjectures include the de Sitter Swampland Conjecture \cite{Obied:2018sgi}, which postulates that there exists a lower bound on the logarithmic slope of scalar field potentials, which severely constrains the model space available for single-field inflation \cite{Kinney:2018nny}. Related are the \textit{refined} de Sitter Swampland Conjecture \cite{Garg:2018reu,Ooguri:2018wrx}, which as the name suggests postulates that there exists a lower bound on the logarithmic slope of scalar field potentials, except when there isn't, and the trans-Planckian Censorship Conjecture, which postulates that sub-Planckian quantum fluctuations must remain quantum mechanical in any consistent UV-complete theory of quantum gravity, conveniently ruling out almost all models of inflation, and suggesting that alternate models for the early universe are preferred \cite{Brandenberger:2021pzy}. This has led to the further development of the Marshland Conjecture by Marsh and Marsh \cite{Marsh:2019lhu}, exploring the ``marshland'' of theories lying on the boundary between the string landscape and the swampland, which is conjectured to be fractal, and predicts a critical number of axionic degrees of freedom $\mathcal{N}_{axion} = 42$. 

This extensive and fashionable body literature, however, neglects to consider fundamental bounds on the number of swampland conjectures themselves. This paper is intended to close this crucial gap in the swampland literature. 

\section{The Swampland Conjecture Bound Conjecture}

The first rule of entropy bounds is: you do not violate entropy bounds. An upper bound on the information content of the observable universe can be obtained from Beckenstein-Hawking entropy a spacetime patch the size of our Hubble volume, which is of order
\begin{equation}
S_H \sim \left(\frac{M_{\mathrm P}}{H_0}\right) \sim 10^{122}.
\end{equation}
Assuming $S_{SW} \sim 10^5$ bits of entropy per paper (a unit of entropy which I propose to name one \textit{ambulance}) there is then a thermodynamic upper bound on the number of papers of
\begin{equation}
N \leq \frac{S_H}{S_{SW}} \sim 10^{117}.
\end{equation}

\begin{figure}[!h]
\begin{center}
\includegraphics[width=\linewidth]{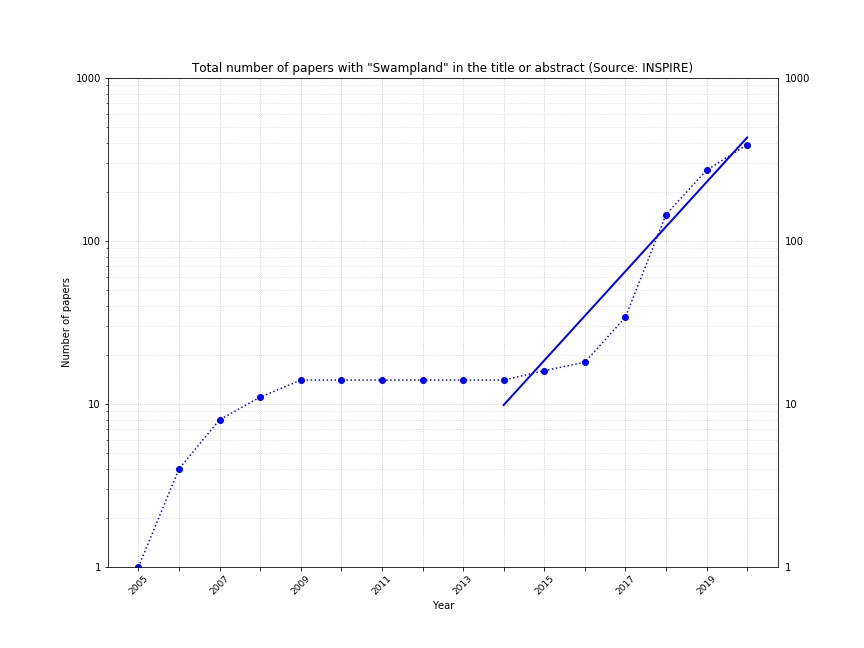}
\caption{The number of papers with ``Swampland'' in the title or abstract, plotted by year. The solid line plots a best-fit doubling time of 1.1 years since the onset of exponential growth in 2014.}
\label{fig:growth}
\end{center}
\end{figure}

While this number of papers is large, it is finite, and current production of swampland papers is exhibiting exponential growth. Figure \ref{fig:growth} shows the total number of INSPIRE papers with the word ``Swampland'' in the title or abstract \cite{Kolb:2021xfn,Brandenberger:2021pzy,Caron-Huot:2021rmr,Jonas:2021xkx,Li:2021gbg,Cecotti:2021cvv,Cvetic:2021lss,vanBeest:2021lhn,Rudelius:2021oaz,Osses:2021snt,Andriot:2021rdy,Park:2021jmi,Barrau:2021vap,Chojnacki:2021fag,Trivedi:2021ivk,Lin:2021ciy,Sadeghi:2020png,Atli:2020dni,Arjona:2020skf,Parikh:2020ggm,Mavromatos:2020kzj,Anderson:2020hux,Raghuram:2020vxm,Oikonomou:2020oex,Calderon-Infante:2020dhm,Dierigl:2020lai,Trivedi:2020xlh,Blumenhagen:2020doa,Mohammadi:2020twg,Winkler:2020ape,Perlmutter:2020buo,Bastian:2020egp,Plauschinn:2020ram,Bonnefoy:2020uef,Cribiori:2020use,Guarino:2020flh,Papageorgiou:2020nex,Baume:2020dqd,Mininno:2020sdb,Bedroya:2020rmd,Palti:2020mwc,Jin:2020vbl,Coriano:2020yso,Li:2020cjj,Eichhorn:2020sbo,Farakos:2020wfc,Herrera:2020mjh,Akrami:2020zfz,Mitchell:2020big,Dine:2020vmr,Kumar:2020jwq,Cvetic:2020kuw,Trivedi:2020wxf,Bedroya:2020rac,Storm:2020gtv,Mavromatos:2020crd,Das:2020xmh,Basile:2020mpt,Angelantonj:2020pyr,Farakos:2020idt,Cicoli:2020noz,Font:2020rsk,Cecotti:2020uek,Buchel:2020xdk,Heidenreich:2020ptx,Brahma:2020eqd,Nortier:2020vge,Elizalde:2020say,Brahma:2020htg,CaboBizet:2020cse,Lanza:2020qmt,Yang:2020jze,Sharma:2020lmm,Mohammadi:2020ake,Ellis:2020nnp,Banerjee:2020xcn,Matsui:2020tyd,Sabir:2020fbi,DeBiasio:2020xkv,Xu:2020nlh,Berera:2020iyn,March-Russell:2020lkq,Conlon:2020wmc,Herraez:2020tih,Bouhmadi-Lopez:2020wve,Emelin:2020buq,Bento:2020fxj,Anchordoqui:2020sqo,Palti:2020tsy,Malek:2020mlk,Gonzalo:2020kke,Low:2020kzy,Sharma:2020wba,Montefalcone:2020vlu,Solomon:2020viz,Abe:2020jgf,Katz:2020ewz,Blumenhagen:2020dea,Benakli:2020pkm,Mohammadi:2020ftb,Gendler:2020dfp,Cecotti:2020rjq,McNamara:2020uza,Odintsov:2020zkl,Bonnefoy:2020fwt,Naskar:2020vkd,Andriot:2020lea,Heisenberg:2020ywd,Luben:2020wim,Palti:2020qlc,Dixit:2020buz,Li:2020tqx,Blumenhagen:2020xpq,Kogai:2020jkq,Junghans:2020acz,Chaudhary:2020yyv,Kamali:2020drm,Brahma:2020cpy,Adhikari:2020xcg,Brandenberger:2020oav,Enriquez-Rojo:2020pqm,Cicoli:2020cfj,Daus:2020vtf,Bravo:2019xdo,Rasheed:2020syk,Wan:2019fxh,Skrzypek:2020dud,Sun:2019obt,Shifman:2019qvx,Fichet:2019owx,Craig:2019zkf,Dudas:2019pls,DallAgata:2019yrr,Bonnefoy:2019nzv,Ferrara:2019tmu,Kim:2019ths,Scalisi:2019gfv,Cai:2019dzj,Anchordoqui:2019amx,Dasgupta:2019rwt,Lin:2019fdk,Saito:2019tkc,Brax:2019rwf,Kehagias:2019iem,Brandenberger:2019jbs,Seo:2019wsh,Li:2019ipk,Channuie:2019vsp,Lin:2019pmj,Dasgupta:2019vjn,Linder:2019caj,Leedom:2019cxx,Laliberte:2019sqc,Geng:2019phi,Banks:2019oiz,Goswami:2019ehb,Vafa:2019evj,Brahma:2019vpl,Berera:2019zdd,Blumenhagen:2019vgj,Sadeghi:2019veh,Grimm:2019ixq,Schmitz:2019uti,Torabian:2019zms,Kamali:2019xnt,Draper:2019utz,Geng:2019zsx,Das:2019hto,Montero:2019ekk,Lee:2019wij,Kehagias:2019akr,Shukla:2019akv,Bedroya:2019snp,Antoniadis:2019rkh,McNamara:2019rup,Shukla:2019dqd,Hebecker:2019csg,Wu:2019xtv,Fichet:2019ugl,Blanco-Pillado:2019tdf,Jones:2019nev,Chakraborty:2019dfh,Dasgupta:2019gcd,Gomez:2019ltc,Yamazaki:2019ahj,Seo:2019mfk,Scherrer:2019dkc,Hardy:2019apu,Wang:2019eym,Baldes:2019tkl,Wang:2019tjh,Han-Yu:2019tmf,McGuigan:2019gdb,Zhai:2019std,Faraggi:2019fap,Damian:2019bkb,Agrawal:2019dlm,Hebecker:2019vyf,Blumenhagen:2019kqm,Lust:2019zwm,Choi:2019mva,Lee:2019skh,Mizuno:2019pcm,Kim:2019vuc,Aragam:2019omo,Heckman:2019bzm,Colgain:2019joh,Benetti:2019smr,Erkinger:2019umg,Rudelius:2019cfh,Rajvanshi:2019wmw,Hamada:2019fmc,Grimm:2019wtx,CaboBizet:2019sku,Brahma:2019iyy,Shirai:2019tgr,Tada:2019amh,vandeBruck:2019vzd,Colgain:2019pck,Lee:2019xtm,Brahma:2019mdd,Font:2019cxq,Marchesano:2019ifh,Geng:2019bnn,Mukhopadhyay:2019cai,Marsh:2019lhu,Gonzalo:2019gjp,Sabir:2019wel,Palti:2019pca,Christodoulidis:2019jsx,Joshi:2019nzi,Brahma:2019kch,Bjorkmo:2019fls,Andriot:2019wrs,Haque:2019prw,Blumenhagen:2019qcg,Park:2019odd,Lynker:2019joa,Junghans:2019azy,Heisenberg:2019qxz,Artymowski:2019vfy,Kamali:2019hgv,Bramberger:2019zez,Bjorkmo:2019aev,Kobakhidze:2019ppv,Kallosh:2019axr,Taylor:2019ots,Cai:2018ebs,Raveri:2018ddi,Abel:2018zyt,Corvilain:2018lgw,Seo:2018abc,Scalisi:2018eaz,Bastero-Gil:2018yen,Gonzalo:2018guu,Hamada:2018qef,Hebecker:2018fln,Buratti:2018xjt,Lin:2018edm,Hertzberg:2018suv,deRham:2018dqm,Kinney:2018kew,Herdeiro:2018hfp,Bonnefoy:2018tcp,Acharya:2018deu,Andriot:2018mav,Banlaki:2018ayh,Klaewer:2018yxi,Emelin:2018igk,Junghans:2018gdb,Conlon:2018vov,Holman:2018inr,Blanco-Pillado:2018xyn,Tosone:2018qei,Ibe:2018ffn,Elizalde:2018dvw,Cheong:2018udx,Thompson:2018ifr,Grimm:2018cpv,Yi:2018dhl,Chiang:2018lqx,Agrawal:2018rcg,Lin:2018rnx,Schimmrigk:2018gch,Park:2018fuj,Blaback:2018hdo,Dvali:2018jhn,Garg:2018zdg,Gautason:2018gln,Olguin-Tejo:2018pfq,Hebecker:2018vxz,Buratti:2018onj,Wang:2018kly,Fukuda:2018haz,Craig:2018yld,Ooguri:2018wrx,Das:2018rpg,Lee:2018spm,Ashoorioon:2018sqb,Odintsov:2018zai,Motaharfar:2018zyb,Kawasaki:2018daf,Hamaguchi:2018vtv,Lin:2018kjm,Matsui:2018xwa,Dimopoulos:2018upl,Bena:2018fqc,Han:2018yrk,Wang:2018duq,Danielsson:2018qpa,Das:2018hqy,Choi:2018rze,Raghuram:2018hjn,Brahma:2018hrd,Murayama:2018lie,Marsh:2018kub,Heisenberg:2018rdu,Reece:2018zvv,Akrami:2018ylq,Cicoli:2018kdo,Dasgupta:2018rtp,Kinney:2018nny,Lee:2018urn,Conlon:2018eyr,Cecotti:2018ufg,Damian:2018tlf,Heisenberg:2018yae,Chiang:2018jdg,Ben-Dayan:2018mhe,Matsui:2018bsy,Roupec:2018mbn,Andriot:2018ept,Colgain:2018wgk,Denef:2018etk,Dias:2018ngv,Rasouli:2018kvy,Kehagias:2018uem,Lehners:2018vgi,Garg:2018reu,Achucarro:2018vey,Hebecker:2018ofv,Branchina:2018xdh,Andriot:2018wzk,Agrawal:2018own,Obied:2018sgi,Andriot:2018tmb,Landete:2018kqf,Heckman:2018jxk,Moritz:2018sui,Blumenhagen:2018hsh,Aldazabal:2018nsj,Blumenhagen:2018nts,Taylor:2018khc,Heidenreich:2018kpg,Cicoli:2018tcq,Halverson:2018vbo,Brennan:2017rbf,Giombi:2017mxl,Montero:2017mdq,Hamada:2017yji,Ibanez:2017oqr,Fisher:2017dbc,Cvetic:2017epq,Klevers:2017aku,Ibanez:2017kvh,Palti:2017elp,Valenzuela:2017bvg,Blumenhagen:2017cxt,Ko:2017iki,Montero:2017yja,Ooguri:2016pdq,Johnson:2016qar,Antonelli:2015oza,Brown:2015iha,Stamnitz:2009ytw,Kumar:2009us,Yaida:2009jz,Fiol:2008gn,Chen:2008gi,Sharpe:2008rd,Huang:2007st,Wu:2007pq,Moura:2007ac,Huang:2007qz,Green:2007zzb,Gasperini:2006as,Kats:2006xp,Ooguri:2006in,Vafa:2005ui}, plotted by year. The solid line shows a best-fit exponential growth curve, with a doubling time of $t_0 = 1.1$ years. Using this rate of growth as a benchmark, we can then estimate a heat-death time of the universe due to the production of swampland conjectures to be of order
\begin{equation}
t \sim 1.1 \times \log_2{\left(10^{117}\right)} \sim 427\ \mathrm{years},
\end{equation}
or in units of a Hubble time,  $t \sim 3 \times 10^{-8}\ H_0^{-1}$. 

We can also compare the upper bound on the number of Swampland conjectures to the dimensionality $\mathcal{N}_{VAC}$ of the string landscape,
\begin{equation}
   \frac{\mathcal{N}_{VAC}}{\mathcal{N}_{SW}} \sim \frac{10^{500}}{10^{117}} \sim 10^{383}. 
\end{equation}
This suggests that fully exploring the properties of the landscape via conjectural methods will require multiple cosmological horizons for basic reasons of computational complexity. An estimate can be obtained from the duality between the quantum complexity of a holographic state and the volume of a Wheeler-Dewitt spacetime patch \cite{Brown:2015bva}, suggesting that the computational complexity required to simulate our current Hubble volume is of order $\mathcal{C} \sim 10^{120}$ quantum gates \cite{Denef:2017cxt}. If we take the ability to fully quantum-simulate our universe as an upper bound on the computational complexity of any given swampland conjecture, the result is that the number of Hubble volumes $\mathcal{N}_H$ required to fully characterize the string landscape via conjectural methods is conservatively bounded from below by
\begin{equation}
    \mathcal{N}_H > \frac{\mathcal{N}_{VAC}}{\mathcal{C} \mathcal{N}_{SW}} \sim 10^{263}.
\end{equation}
As a meaningless comparison, this is more than $10^{90}$ times as large as the number of legal positions on a standard 19x19 Go board. Which is a lot. Taking the number of atoms in our Hubble volume to be of order $\mathcal{N}_A \sim 10^{80}$, if every atom in the universe were to encode a complete solution to the game of Go, the corresponding multiplicity would still be more than 10 orders of magnitude smaller than the lower bound $\mathcal{N}_H$. 

\section{Conclusions}

So-called ``Swampland'' conjectures aim to constrain the properties of any UV-complete theory such as string theory by the corresponding properties of the low-energy effective limit of the theory. A number of such conjectures have been put forward, such as the Weak Gravity Conjecture, the de Sitter Swampland Conjecture, the Refined de Sitter Swampland Conjecture, the trans-Planckian Censorship Conjecture, and the Marshland Conjecture.  In this paper, I consider fundamental bounds on the entropy generated by these conjectures themselves. I conjecture that an upper bound on the entropy contained in conjectures is given by the Beckenstein-Hawking entropy of the current cosmological horizon, which I call the \textit{Swampland Conjecture Bound Conjecture} (SCBC). Under the assumption of order 100 kilobits of entropy per conjecture, this results on an upper bound on the number of swampland conjectures of order $\mathcal{N} \sim 10^{117}$. Note that this bound is not strongly sensitive to the \textit{ansatz} on the entropy per conjecture, but will be of order $10^{100}$ even if the conjecture itself contains almost no information, for example an emoji or an irate tweet about Elon Musk. At the current exponential growth rate of papers on the Swampland, the SCBC can be projected to be saturated on a timescale of a few hundred years. This cosmic-scale resource requirement suggests that humanity's h-index critically depends on our transition into a spacefaring civilization in the near future. 

I further compare the limit from the Swampland Conjecture Bound Conjecture to the computational complexity of the string landscape itself, consisting of order $\mathcal{N}_{VAC} \sim 10^{500}$ independent possible low-energy completions in the form of geometrically distinct compactifications of the 10- or 11-dimensional manifold of string or $\mathcal{M}$-theory or whatever into a lower-dimensional geometry. The level of complexity represented by the problem of string compactification is daunting, even on cosmic scales. Using bounds on computational complexity motivated by an AdS/CFT duality between a boundary holographic state and spacetime volume, I place a conservative lower bound on the number of Hubble volumes which must be driven to heat death to fully explore the string landscape by means of making guesses about its properties to be of order $\mathcal{N}_H > 10^{263}$, give or take a few dozen orders of magnitude. 

\begin{acknowledgments}
I am grateful to Thomas Van Riet for wisely refusing to have useful discussions. This work is entirely unrelated to any past, present, or future National Science Foundation funding. No animals were harmed in the production of this paper. 
\end{acknowledgments}

\bibliography{SCBC.bib}

%
%

\end{document}